\documentclass[12pt]{article}
\usepackage{graphicx}
\usepackage{amsmath,amssymb}
\usepackage[utf8]{inputenc}
\usepackage{a4}
\def\eps{\varepsilon}

\def \RR {{\mathbb R}}

\def \ZZ {{\mathbb Z}}

\def\CAR{{\rm CAR}}  
\def\omit#1{}

\def\be{\begin{equation}} \def\beno{\begin{equation}\notag} \def\ee{\end{equation}}
\def\bea{\begin{eqnarray}}  \def\eea{\end{eqnarray}} 
\def\ba{\begin{array}} \def\ea{\end{array}}
 \def\inv{^{-1}}

\def\wh{\widehat} 
\def\eins{\mathbf 1} 
\def\wick#1{{}:\!#1\!:{}}

\newcommand{\eref}[1]{(\ref{#1})}

\begin{document}
\title{Multilocal Fermionization} 

\author{Karl-Henning Rehren$^{1,2}$, Gennaro Tedesco$^{1}$}
\maketitle

\begin{center}
\footnotesize
$^1$ Institut f\"ur Theoretische Physik, Universit\"at G\"ottingen, \\
Friedrich-Hund-Platz 1, 37077 G\"ottingen, Germany \\[2mm]
$^2$ Courant Research Centre ``Higher Order Structures in
Mathematics'', \\ 
  Universit\"at G\"ottingen, Bunsenstr.\ 3--5, 37073 G\"ottingen, Germany
\end{center}

\begin{abstract}
We present a simple isomorphism between the algebra of one real chiral
Fermi field and the algebra of $n$ real chiral Fermi fields. The
isomorphism preserves the vacuum state. This is possible by a
``change of localization'', and gives rise to new multilocal
symmetries generated by the corresponding multilocal current and
stress-energy tensor. The result gives a common underlying explanation
of several remarkable recent results on the representation of the free
Bose field in terms of free Fermi fields \cite{A1,A2}, and on the modular
theory of the free Fermi algebra in disjoint intervals
\cite{CH,LMR}. 
\end{abstract}



\section{Introduction}
\setcounter{equation}{0}

It is well known that there is an algebraic isomorphism (the ``split
isomorphism'') $A(x)B(y)\mapsto A(x)\otimes^t B(y)$,
$x\in O_1$, $y\in O_2$ between the fields of a given QFT
localized in two spacetime regions $O_i$, and two independent copies 
of the same fields localized in the same regions, as long as the
regions are sufficiently well spacelike separated from each other.\footnote{ 
The graded tensor product $A\otimes^tB$ is the true tensor product
$A\otimes B$ if $A$ is a Bose field, and a twisted tensor product
$A\otimes (-1)^FB$ if $A$ is a Fermi field, such that $\eins\otimes^t B$ 
and $A\otimes^t\eins$ anticommute when $A$ and $B$ are both Fermi fields.}  

Naively, this is true because $A$ (anti)commute with $B$ by graded
locality, while $A\otimes^t\eins$ (anti)commute with
$\eins\otimes^tB$ by construction, so that all algebraic relations
are preserved. However, this isomorphism does not preserve the vacuum
state, because it eliminates all correlations between fields in $O_1$
and fields in $O_2$: $\omega(A(x)\otimes^t B(y)) = 
\omega(A(x))\cdot\omega(B(y))\neq\omega(A(x)B(y))$. Moreover, it
cannot be defined globally, because the restriction to spacelike
separated regions is essential in the argument. 

We present here a new isomorphism between the algebra of one real chiral
Fermi field, and the algebra of a complex = two real chiral Fermi
fields (or actually of any number $n$ of real Fermi fields), which
preserves the vacuum state and is globally defined. The price to pay
is a change of the notion of localization.   

This means in particular, that the isomorphism does not intertwine the
respective conformal transformations. We shall discuss the relation
between the stress-energy tensors of the real Fermi field
and of the complex Fermi field, and features of the ``embedded''
diffeomorphisms generated by either SET acting on the other field. We
shall also present the bilocal gauge transformations of the real Fermi
field generated by the current of the complex Fermi field.

This isomorphism provides a very simple understanding of two recent
remarkable results: 

The first is the fact \cite{A2} that besides the standard
fermionization formula $j(x)=\wick{\phi^*(x)\phi(x)}$ of the free
chiral Bose current in terms of a complex Fermi field, one can obtain
the same current as a bilocal Wick product of a real Fermi field at two
different points: $j(x)\sim \wick{\psi(x_1)\psi(x_2)}$, where $x_2=-1/x_1$
are the two solutions of $q(x_i)\equiv 2x_i/(1-x_i^2)=x$. (The
differential algebra underlying this relation was described much
earlier in \cite{DJKM} in the context of hierarchies of integrable
systems.)  

The isomorphism generalizes to an isomorphism of one real Fermi
field with any number $n$ of real Fermi fields (at the price of an $n$-fold
localization). Since non-abelian current algebras can be constructed
as certain subalgebras of $n$-free-Fermi quantum field theories, it
would be interesting to study the resulting multilocal embedding into
the theory of a single free Fermi field.   

The second remarkable recent result is the discovery \cite{CH} (see also
\cite{LMR}) that the modular automorphism group of the local von
Neumann algebra of a free Fermi theory in a union of disjoint
intervals in the vacuum state acts ``almost'' geometric. The geometric
part of the ``modular dynamics'' is given by the pullback of the
dilation group under a rational universalizing function that maps each
of the intervals onto $\RR_+$ and each component of the complement
onto $\RR_-$; in addition, there occurs a ``mixing'' among the fields at related
points in each of the intervals.

Under the present isomorphism, the $n$ Fermi fields at a point $X$
are mapped to position-dependent linear combinations of a single Fermi
field at the $n$ pre-images $x_k$ of $X$. Because the isomorphism
preserves the vacuum state, it intertwines the respective modular 
automorphisms. Therefore, the modular mixing of one Fermi field in $n$
intervals is naturally explained from the well-known modular
automorphism group of $n$ independent Fermi fields in a single
interval, which is just the subgroup of the M\"obius group preserving
the interval \cite{FG,BGL}.  

\medskip

{\bf Remark:} With the term ``fermionization'', we refer to the
representation of Bose fields in terms of Fermi fields, as the
opposite of ``bosonization'', the representation of Fermi fields in
terms of Bose fields \cite{M,CRW,S}.

\section{The starting point}
\setcounter{equation}{0}

Let $\psi$ be the real chiral Fermi field: $\psi(x)^*=\psi(x)$, with CAR 
\be \label{CAR-r}
\{\psi(x),\psi(y)\}=2\pi\,\delta(x-y),\qquad (x,y\in\RR)
\ee
and vacuum two-point function 
\beno
\omega\big(\psi(x)\psi(y)\big)
=\frac{-i}{x-y}\equiv\lim_{\eps\searrow0}\frac{-i}{x-y-i\eps}.
\ee
Let $\phi$ be a complex chiral Fermi field: $\phi(x)^*=\phi^*(x)$, 
with CAR
\be \label{CAR-c}
\{\phi(x),\phi^*(y)\}=\{\phi^*(x),\phi(y)\}=2\pi\,\delta(x-y),\quad
\{\phi,\phi\}=\{\phi^*,\phi^*\}=0 \qquad
\ee
and vacuum two-point function 
\beno
\omega\big(\phi(x)\phi^*(y)\big) = \omega\big(\phi^*(x)\phi(y)\big) 
=\lim_{\eps\searrow0}\frac{-i}{x-y-i\eps},\quad
\omega\big(\phi\phi\big)=\omega\big(\phi^*\phi^*\big)=0.
\ee
The complex Fermi field can be decomposed into two anti-commuting
real Fermi fields:
\beno
\phi(x)=\big(\psi^{(1)}(x)+i\psi^{(2)}(x)\big)/\sqrt2.
\ee
We write $\CAR$ and $\CAR^n$ for the algebras generated by one, resp.\
$n$ real Fermi fields.

Let us also introduce the ``compact picture'', using the  Cayley transformation
\be \label{Cayley}
z=\frac{1+ix}{1-ix}\in S^1\setminus\{-1\}
\ee
and the definition
\be \label{compact}
\wh\psi(z):= \Big(-i\frac {d\,z}{dx}\Big)^{-\frac 12}\psi(x)\equiv
\frac {1-ix}{\sqrt2}\,\psi(x)
\ee
and likewise for all other Fermi fields\footnote{The present choice of
  the branch of the square root in the transformation law determines
  the branch of similar square roots throughout.}. Then, for the real field, one
has $\wh\psi(z)^*=z\wh\psi(z)$, and for the complex field
$\wh\phi(z)^*=z\wh{\phi^*}(z)$. The non-vanishing two-point functions are 
\beno
\omega\big(\wh\psi(z)\wh\psi(w)\big)=\omega\big(\wh\phi(z)\wh{\phi^*}(w)\big)=\omega\big(\wh{\phi^*}(z)\wh\phi(w)\big)=\frac
1{z-w}\equiv\lim_{\lambda\nearrow1}\frac 1{z-\lambda w},
\ee
and the CAR is given in terms of $\lim_{\lambda\nearrow1}(\frac 1{z-\lambda w} + \frac
1{w-\lambda z}) = 2\pi\delta(z,w)\equiv\frac
{2\pi}z\delta(\varphi-\vartheta)$ for $z=e^{i\varphi}$, $w=e^{i\vartheta}$, $\varphi,\vartheta\in(-\pi,\pi)$. 

The ``compact picture'' CAR algebras on the cut circle
$S^1\setminus\{-1\}$  are just reparametrizations of the CAR algebras 
on $\RR$. Their extension to the full circle $S^1$ depends 
on the representation. The CAR algebra on $\RR$ possesses two faithful
representations: the vacuum (or Neveu-Schwarz) representation and the
Ramond representation \cite{Fu}. The former extends periodically to $S^1$, 
while the latter extends anti-periodically to $S^1$, i.e., it extends
to the two-fold covering of $S^1$. 

Our starting point (Prop.\ 1) is an isomorphism between the complex
Fermi algebra $\CAR^2(S^1)$ and the real Fermi algebra $\CAR(S^1)$ in
the vacuum representation. We shall give two simple proofs. The
simplest one is in the global setting of the Fourier modes. We also
give another, local proof, because our main interest in the rest of
the paper lies in the local properties of the resulting new symmetries
of the real Fermi field, which can be understood by this
isomorphism. These symmetries 
are in a controlled way nonlocal (``bilocal''), If they were
completely nonlocal, there would be little interest; we believe,
however, that the bilocal symmetries are of some physical
relevance. This will be elaborated in Sect.\ 4, where contact with the
modular theory of the real Fermi field is made.

Due to the different global behaviour, the following proposition holds
only in the vacuum representation. It entails bilocal fermionization
formulae for the current and the stress-energy tensor in the vacuum
representation (Cor.\ 1 and 3). Interestingly enough, even if there is
no analogue of the proposition in the Ramond sector, the bilocal
fermionizations persists. This aspect will be discussed in Sect.\ 5.      

\medskip 

{\bf Proposition 1:} {\sl Let $\phi$ and $\psi$ stand for the complex
  and real Fermi fields in the vacuum representation. The linear map 
\bea\notag
\beta:\wh\phi(z^2)\mapsto \frac12\big(\wh\psi(z)+\wh\psi(-z)\big), 
\\ \label{beta}
\wh{\phi^*}(z^2)\mapsto
\frac1{2z}\big(\wh\psi(z)-\wh\psi(-z)\big)
\eea
for $z\in S^1$, induces an isomorphism
$\beta:\CAR^2(S^1) \to \CAR(S^1)$ of CAR algebras, which preserves
the vacuum state.}   

\medskip 

Note that the map is well-defined because the right-hand sides are
invariant under $z\leftrightarrow-z$.

{\em Proof:} The simplest proof proceeds by looking at the Fourier
modes of the real and complex free Fermi fields,
\be \notag
\wh\psi(z)=\sum_{n\in\ZZ+\frac12}\psi_n z^{-n-\frac12},
\quad\wh\phi(z)=\sum_{n\in\ZZ+\frac12}\phi_n z^{-n-\frac12}, \quad
\wh{\phi^*}(z)=\sum_{n\in\ZZ+\frac12}(\phi^*)_n z^{-n-\frac12}, \quad
\ee
where $\psi_n=(\psi_{-n})^*$ and $(\phi^*)_n=(\phi_{-n})^*$ satisfy
the CAR $\{\psi_n,\psi_m\}=\delta_{n+m,0}$,
$\{\phi_n,\phi^*_m\}=\delta_{n+m,0}$. In terms of these modes, the
isomorphism is given by
\beno
\beta\big(\wh\phi(z^2)\big) = \sum_{n\in\ZZ+\frac12} \psi_{2n+\frac12}
(z^2)^{-n-\frac12},\quad \beta\big(\wh{\phi^*}(z^2)\big) =
\sum_{n\in\ZZ+\frac12} \psi_{2n-\frac12} (z^2)^{-n-\frac12}.
\ee
This is just a renumbering of the generators
\be
\phi_n\mapsto\psi_{2n+\frac12},\qquad(\phi^*)_n\mapsto\psi_{2n-\frac12}
\qquad \big(n\in\ZZ+\frac12\big),
\ee
such that the conjugate complex fields $\beta(\wh\phi)$, 
$\beta(\wh{\phi^*})$ involve the Fourier modes $\psi_m$ of the real
Fermi field with $m\in 2\ZZ-\frac12$, $m\in2\ZZ+\frac12$, respectively.

It is then obvious that the CAR and adjoint relations are
preserved. The vacuum state is preserved because the vacuum is the
unique state which is annihilated by all $\psi_n$ ($n>0$), and by all
$\phi_n,\phi^*_n$ ($n>0$), respectively. This already proves the statement.

We wish, however, to emphasize the local aspects of the isomorphism. 
Therefore, we also indicate the proof in the setting of local quantum
field theory.  

One may directly compute the anti-commutators
\beno
\{\beta(\wh{\phi^*}(z^2)),\beta(\wh{\phi}(w^2))\}=2\pi\delta(z^2,w^2)=\frac\pi
z\big(\delta(z,w)+\delta(z,-w)\big)
\ee
etc. It is however sufficient and in fact much easier, to verify the
equality of the vacuum two-point functions
\beno
\omega\big(\beta(\wh{\phi^*}(z^2))\beta(\wh{\phi}(w^2))\big)=\frac1{z^2-w^2}
\ee
etc, which is easily done by a direct computation. By Wick's theorem,
this equality then extends to all $n$-point functions. The CAR follow
from these correlation functions. 

The adjoint relation 
\beno
\beta\big(\wh\phi(z^2)\big)^*=z^2\beta\big(\wh{\phi^*}(z^2)\big)
\ee
is immediate. Thus $\beta$ is a *-homomorphism. 

Finally, $\beta$ is an isomorphism because it has an inverse:
\be \label{betainv}
\beta\inv\big(\wh\psi(z)\big) = \wh\phi(z^2)+z\wh{\phi^*}(z^2).
\ee

This proves the proposition. \hfill $\square$

\bigskip

Going back to the non-compact picture, the isomorphism can be written as
\bea\notag 
\phi(q(x))\mapsto\frac1{q(x)}\Big[\frac{x\psi(x)}{1-ix}+\frac{i\psi(-\frac
  1x)}{1+ix}\Big],
\\ \label{beta-phix}
\phi^*(q(x))\mapsto\frac1{q(x)}\Big[\frac{x\psi(x)}{1+ix}-\frac{i\psi(-\frac
  1x)}{1-ix}\Big],
\eea
where $q(x)=\frac{2x}{1-x^2}$ is the map $(-1,1)\to\RR$ corresponding,
under the Cayley transformation, to the square map on $z\in S^1$.
The inverse reads
\be \label{betainv-kx}
\beta\inv(\psi(x))
=\frac{1+ix}{1-x^2}\;\phi(q(x))+\frac{1-ix}{1-x^2}\;\phi^*(q(x)).
\ee

\bigskip

Equipped with this isomorphism, we shall discuss several
implications in the sequel of this paper:

\begin{enumerate}

\item Under this isomorphism, the standard local fermionization
formula for the current in terms of the complex CAR algebra
\be
\wh j(z)=\wick{\wh{\phi^*}(z)\wh\phi(z)} 
=i\wick{\wh{\psi^{(1)}}(z)\wh{\psi^{(2)}}(z)}
\ee
turns into the new bilocal fermionization formula \cite{A2} embedding
the current into the real CAR algebra
\be
2z\cdot \beta\big(\wh j(z^2)\big)=\wick{\wh{\psi}(z)\wh{\psi}(-z)},
\ee
(and a similar formula for the stress-energy tensor). We shall present
the ``bilocal gauge transformations'' generated by this ``embedded''
current in Sect.\ 3. 

\item The embedded stress-energy tensor consists of two pieces: the
stress-energy tensor of the real Fermi theory, and an embedded
current. Consequently, the embedded diffeomorphisms consist of a
geometric flow and a bilocal gauge transformation. This is in
particular true for the embedded M\"obius transformations, which are
known to arise as modular automorphisms in the vacuum state
\cite{FG,BGL}. Therefore, the ``modular mixing'' \cite{CH,LMR} of the
Fermi field in multiple intervals finds its explanation as embedded
M\"obius transformations. Details of this will be presented in Sect.\
4. 

\item There is an analogous isomorphism between the real Fermi field
  in the Ramond representation and a pair of real Fermi fields, one in
  the Ramond, one in the vacuum representation. This yields another
  embedding of the current algebra into the Ramond representation of
  the real Fermi field, which turns out to be the ``twisted''
  representation of the current \cite{Fr}. The latter is defined by
  the quasifree state\footnote{A state on a CAR or CCR algebra is
    called quasifree if it obeys Wick's theorem, i.e., all higher
    correlation functions are sums of products of 2-point functions.} 
  with two-point function   
\bea \label{twisted}
\omega_t\big(\wh j(z)\wh
j(w)\big)&=&\frac{w+z}{2\sqrt{wz}}\cdot\frac1{(w-z)^2} \\ \notag
\omega_t\big(j(x)j(y)\big) &=&
\frac{1+xy}{\sqrt{(1+x^2)(1+y^2)}}\cdot\frac{-1}{(x-y)^2}, 
\eea
in the compact and non-compact picture, respectively. The twisted
representation has lowest conformal   energy $L_0=\frac1{16}$. Some  
aspects of this bilocal current will be discussed in Sect.\ 5.
\end{enumerate}

\medskip

To conclude this section, we notice that Prop.\ 1 generalizes to an
isomorphism between one real Fermi field and any number $n$ of real
Fermi fields, exploiting the map $z\mapsto z^n$. Let us combine $n$
real Fermi fields into complex Fermi fields $\phi^{(k)}$
($k=1,\dots, n$) such that $(\phi^{(k)}(x))^*=\phi^{(n+1-k)}(x)$. If
$n$ is odd, $\phi^{(\frac{n+1}2)}$ is real.

\medskip\newpage

{\bf Proposition 2:} {\sl The linear map   
\bea\label{beta-n}
\beta:\wh{\phi^{(k)}}(z^n)\mapsto \frac {z^{1-k}}n \sum_{j=0}^{n-1}
\omega^{(1-k)j} \cdot \wh\psi(\omega^jz) \qquad (z\in S^1,\,k=1,\dots,n),
\eea
where $\omega=e^{\frac{2\pi i}n}$, induces an isomorphism
$\beta:\CAR^n(S^1) \to \CAR(S^1)$ of CAR algebras, which preserves
the vacuum state.}

That this is a vacuum-preserving isomorphism, can again be verified as
in the proof of Prop.\ 1, either by a direct computation of the vacuum
expectation values, from which also the anti-commutation relations
follow, or by noting that $\beta\big(\psi^{(k)}(z^n)\big) =
\sum_{\nu\in\ZZ+\frac12}\psi_{\frac12-k+(\nu+\frac12)n} (z^n)^{-\nu-\frac12}$.

\medskip

In particular, since non-abelian current algebras can be embedded into
free Fermi theories with sufficiently many free Fermi fields, one
obtains representations of all these theories in the Fock space of a
single free Fermi field.

\section{Symmetries}
\setcounter{equation}{0}

Throughout this section, we shall work exclusively in the compact
picture, which drastically simplifies most formulae. The passage to
the non-compact picture can always be made by the transformation laws
$A(x)=\big(\frac{\sqrt2}{1-ix}\big)^{2d}\wh A(z)$ for fields of
dimension $d$. We also suppress the symbol $\wh\cdot$ in this section.

\subsection{Gauge transformations}

The complex free Fermi field algebra is invariant under gauge
transformations. The latter can be implemented by unitary Weyl operators:
\bea\notag
\alpha_f\big(\phi(z)\big) \equiv e^{-if(z)}\phi(z)&=&W(f)\phi(z)W(f)^*,\\ \notag
\alpha_f\big(\phi^*(z)\big) \equiv e^{if(z)}\phi^*(z) &=&W(f)\phi^*(z)W(f)^*,
\eea
where $f:S^1\to\RR$ is a smooth periodic function. In the vacuum
representation, the Weyl operators are given by $W(f)=e^{ij(f)}$,
$j(f)= \oint f(z)j(z)dz/i$. The current $j(z)$ is contained in the
Wick algebra of the complex Fermi field as 
\be \label{standard}
j(z)=\wick{\phi^*(z)\phi(z)}=i\wick{\psi^{(1)}(z)\psi^{(2)}(z)}.
\ee
It satisfies the CCR algebra 
\be \label{CCR}
[j(f),j(g)]=2\pi i\,\oint f(z)\partial_z g(z) dz,
\ee
and its vacuum 2-point function is
\beno
\omega\big(j(z)j(w)\big)= \frac{1}{(z-w)^2}.
\ee

The current algebra defines a free Bose quantum field of its own. The
CCR algebra possesses automorphisms 
\beno
\rho_\lambda(j(z))=j(z)+\lambda(z),
\ee
where $z\lambda(z)$ is a smooth real function. It is
well-known, that states $\omega_\lambda=\omega_0\circ\rho_\lambda$ are
charged states of 
charge $q=\frac1{2\pi}\int\lambda(z)dz$, and these states are ground
states for the conformal Hamiltonian $L_0$ if $\lambda(z)=\frac
{q}{z}$. In this case, we denote $\rho_\lambda=\rho^q$:
\be \label{rho}
\rho^q(j(z))=j(z) + \frac qz.
\ee

Because the Wick product is defined by the subtraction of vacuum
expectation values, and the isomorphism $\beta$ respects the vacuum
state, the latter extends to the Wick algebra of the complex Fermi
field. In particular, we can ``embed'' the current into the real Fermi
algebra. A straightforward computation gives rise to   

\medskip

{\bf Corollary 1:} {\sl The embedded current is
\bea 
\label{nlf}
\beta\big(j(z^2)\big) =
\frac{1}{2z}\wick{\psi(z)\psi(-z)}.
\eea
}

\smallskip

We may also write this in terms of Fourier modes
\beno 
j(z)=\sum_{n\in\ZZ}j_nz^{-n-1} ,
\ee 
where $j_n^*=j_{-n}$ satisfy the CCR
$[j_m,j_n]=m\delta_{m+n,0}$. Then 
\beno
\beta\big(j(z^2)\big)=\frac 1{2z}
\sum_{m,k\in\ZZ+\frac12}\wick{\psi_m\psi_k}(-1)^{-k-\frac12}z^{-m-k-1}.
\ee
If $N\in\ZZ$ is odd, the sum 
$\sum_{k\in\ZZ+\frac12}\wick{\psi_{N-k}\psi_k}(-1)^{-k-\frac12} = 
\sum_{k\in\ZZ+\frac12}\wick{\psi_k\psi_{N-k}}(-1)^{k-N-\frac12}$
vanishes by virtue of the CAR, because
$(-1)^{k-N-\frac12}=(-1)^{-k-\frac12}$. Thus, the sum contains only
odd powers $z^{-N-1}=z^{-2n-1}$, and the expansion in $z^2$ yields
\beno
\beta(j_n) = 
\sum_{\nu=0}^\infty(-1)^{n+\nu+1}\psi_{n-\nu-\frac12}\psi_{n+\nu+\frac12}.
\ee

\medskip

The following corollary gives the action of the embedded gauge
transformations on the real free Fermi field: 

\medskip

{\bf Corollary 2:} {\sl The embedded gauge transformations
  $\beta(W(f)) A \beta(W(f))^* = \beta\circ\alpha_f\circ\beta\inv(A)$ act on the real
  Fermi field by
\beno
\beta(W(f))\psi(z)\beta(W(f))^* =
\cos f(z^2)\cdot \psi(z)+\sin f(z^2) \cdot \psi(-z) .
\ee
}

\smallskip
 
The characteristic feature is the bilocal ``mixing'' of $\psi(z)$ and
$\psi(-z)$, reflecting the nonlocality of the isomorphism $\beta$.

\subsection{Diffeomorphisms}

The real and complex Fermi field algebras are invariant under
diffeomorphisms $\gamma:S^1\to S^1$, implemented by unitary operators 
\beno
V(\gamma)\psi(z)V(\gamma)^*= \sqrt{\gamma'(z)}\cdot\psi(\gamma(z)).
\ee
The unitary implementers of one-parameter groups of diffeomorphisms
are given by  
\beno
V(\gamma_t)=e^{itT(f)}\equiv e^{it\oint f(z)T(z)dz}
\ee
where $if(z)/z\in\RR$. The infinitesimal diffeomorphisms given by
derivations  
\beno
\delta_f(\psi(z))\equiv i[T(f),\psi(z)] = \big(-f(z)\partial_z -\frac12f'(z)\big)\psi(z)
\ee
integrate to finite diffeomorphisms via
$\partial_t\gamma_t(z)=-f(\gamma_t(z))$. 

The stress-energy tensor of the real Fermi field is 
\beno
T^{c=\frac12}(z)= \frac{-1}{4\pi}\wick{\psi\partial\psi}(z) = 
\frac{-1}{8\pi}\wick{\psi\stackrel{\leftrightarrow}{\partial}\psi}(z) ,
\ee
the one of the complex Fermi field is
\beno
T^{c=1}(z)= 
\frac{-1}{4\pi}\wick{\psi^{(1)}\partial\psi^{(1)}}(z)
+\frac{-1}{4\pi}\wick{\psi^{(2)}\partial\psi^{(2)}}(z)
=\frac{-1}{4\pi}\wick{\phi^*\stackrel{\leftrightarrow}{\partial}\phi}(z).
\ee

The current algebra is also diffeomorphism invariant, 
\beno
V(\gamma)j(z)V(\gamma)^*= \gamma'(z)\cdot j(\gamma(z)),
\ee
with the stress-energy tensor given by 
\be \label{Tj2}
T^{\rm curr}(z)= \frac 1{4\pi}\wick{j^2}(z).
\ee
If the current is expressed by the complex Fermi field as above,
then this coincides with $T^{c=1}$:
\be \label{standard-T}
\frac 1{4\pi}\wick{j^2}(z)=
\frac{-1}{4\pi}\wick{\phi^*\stackrel{\leftrightarrow}{\partial}\phi}(z).
\ee

\medskip

Our isomorphism $\beta$ embeds the stress-energy tensor $T^{c=1}$
into the real free Fermi theory. Again, the computation is
straightforward. The result is

\medskip

{\bf Corollary 3:} {\sl The embedded complex stress-energy tensor is
\bea 
\label{betaT-j}
\beta\big(T^{c=1}(z^2)\big) &=& -\frac{1}{8\pi z^2}\;\beta\big(j(z^2)\big) +\frac1{4z^2}\Big(T^{c=\frac12}(z)+T^{c=\frac12}(-z)\Big).\eea
The embedded infinitesimal diffeomorphisms $[i\beta(T^{c=1}(f)),A] =
\beta\circ\delta_f\circ\beta\inv(A)$ act on the real
Fermi field by
\be i\big[\beta\big(T^{c=1}(f)\big),\psi(z)\big] = \Big(-\frac1{2z}f(z^2)\partial_z -\frac12f\,{}'(z^2)\Big)\psi(z)  + \frac1{4z^2}f(z^2)\big(\psi(z) - \psi(-z)\big).\ee
}

Again, we have a a mixing of $\psi(z)$ and $\psi(-z)$, due to
the first contribution in \eref{betaT-j}, on top of a geometric flow due to
the second term. 

The contribution from the current in \eref{betaT-j} can be removed by
composition with a charged automorphism of the current algebra. Thus,
the relation \eref{betaT-j} may also be written as  
\bea 
\label{betaT} 
\beta\circ\rho^{\frac14}\big(T^{c=1}(z^2)\big) &=& 
\frac1{4z^2}\Big(T^{c=\frac12}(z)+T^{c=\frac12}(-z)\Big) + \frac{1}{64\pi z^4}.
\eea

\smallskip 

One may embed conversely the stress-energy tensor of the real Fermi field
into the algebra of the complex Fermi field. Because of the form
\eref{betainv} of the inverse isomorphism $\beta\inv$, this will
involve terms $\wick{\phi\partial\phi}$ and its conjugate, so that the
embedded diffeomorphisms of the real Fermi field will generate
transformations involving the charge conjugation 
$\phi\leftrightarrow\phi^*$.

\medskip

{\bf Corollary 4:} {\sl The embedded real stress-energy tensor is
\be \label{betainvT}
\beta\inv\big(T^{c=\frac12}(z)\big) = -2z^2 T^{c=1}(z^2) +\frac
1{4\pi} j(z^2) -\frac z{2\pi} \big(\wick{\phi\partial\phi}(z^2)
+ z^2\wick{\phi^*\partial\phi^*}(z^2)\big).
\ee
The embedded
infinitesimal diffeomorphisms
$[i\beta\inv(T^{c=\frac12}(f)),A] =
\beta\inv\circ\delta_f\circ\beta(A)$ act on the complex Fermi field by
\bea \notag
 i\big[\beta\inv\big(T^{c=\frac12}(f)\big),\phi(z^2)\big]
= -f_-(z^2)\phi'(z^2) 
-\frac12f'_-(z^2)\phi(z^2) + \frac1{4z^2}f_-(z^2)\phi(z^2) 
\\ 
 -z^2f_+(z^2)\phi^*{}'(z^2) -\frac{z^2}2
f_+'(z^2)\phi^*(z^2) -\frac12 f_+(z^2)\phi^*(z^2) \qquad
\eea
where $f_-(z^2)=z(f(z)-f(-z))$, $f_+(z^2)=f(z)+f(-z)$.
}

\medskip

Let us also introduce the Fourier modes, $\wh T(z)=\sum_{n\in\ZZ}
T_nz^{-n-2}$. The standard Virasoro generators are related to the Fourier modes
$T_n$ by $L_n=2\pi T_n$. Then the relations \eref{betaT-j} and \eref{betaT} are equivalent to 
\be \label{Tmodes}
\beta(L^{c=1}_n)= -\frac14 \beta(j_n) + \frac12 L^{c=\frac12}_{2n},
\qquad \beta\circ\rho^{\frac14}(L^{c=1}_n) =
\frac12 L^{c=\frac12}_{2n} + \frac1{32}\,\delta_{n,0},
\ee
where $\rho^q$ is the charged automorphism \eref{rho} extended to the
stress-energy tensor contained in the Wick algebra of the current.

The latter expression is well known as the embedding of the
infinitesimal ``2-diffeomorphisms'' (i.e., the diffeomorphisms of $z$
induced from those of the variable $z^2$) as a subalgebra of the
Virasoro algebra. Note that the application of $\rho^{\frac14}$ to 
the embedded current \eref{nlf} exactly ``undoes'' the subtraction of
the vacuum expectation in the definition as a Wick product. Indeed,
the subtraction is not necessary, because the points are split
anyway. But the ``unsubtracted'' current has a nonvanishing 
vacuum expectation value, i.e., the corresponding state is a state of
charge $q=\frac14$.

In contrast, the first expression in \eref{Tmodes} involves also the
modes of the current, which may in turn be expressed as before by the
real Fermi modes. The present form emphasizes that the embedded
diffeomorphisms come along with embedded gauge
transformation, i.e., a mixing of $\psi(z)$ and $\psi(-z)$, as
described by Cor.\ 2. 

\section{Modular theory}
\setcounter{equation}{0}

This brings us to the second issue, which finds a simple explanation
by Prop.\ 1.  

``Modular automorphisms'' are a one-parameter group of automorphisms
of any von Neumann algebra $M$ in a faithful normal state $\omega$,
canonically associated with the pair $(M,\omega)$. It is of
particular interest in quantum physics, because the state is
automatically a thermal equilibrium state for the time evolution given
by the modular automorphisms (``modular dynamics'') \cite{HHW}, and
because the modular automorphisms of local algebras for certain simple
spacetime regions coincide with Lorentz or conformal transformations
\cite{BW,FG,BGL}. The latter fact has been exploited \cite{B,GLW,KW} 
to show that the full content of a quantum field theory can be encoded in the
vacuum state and a small finite number of von Neumann algebras ``in
suitable modular position'', such that their modular groups generate
all spacetime symmetries, and all local algebras can be identified as
intersections of the transforms of the initial algebras.   

In the case of chiral conformal theories, the modular automorphisms
for the algebras of observables in an interval $I$ coincide with the
one-parameter subgroup of M\"obius transformations that preserve the
interval \cite{FG,BGL}, namely the dilations 
$\lambda_{-2\pi t}:x\mapsto e^{-2\pi t}x$ 
if $I=\RR_+$ in the non-compact picture: 
\be \label{mod}
\sigma_t\big(\psi(x)\big) = U(\lambda_{-2\pi t})\psi(x)U(\lambda_{-2\pi t})^* = 
e^{-\pi t}\psi(e^{-2\pi t}x) \qquad
(I=\RR_+).
\ee
For every other interval, the modular automorphisms are obtained by
conjugation with a M\"obius transformation that maps $I$ onto
$\RR_+$. Note that M\"obius transformations intertwine the modular
automorphisms, because they preserve the vacuum state. 

\smallskip

The modular automorphisms of the von Neumann algebra of the Fermi
field in multiple intervals in the vacuum state was found in \cite{CH},
and elaborated in \cite{LMR}. It is given by a combination of a
geometric flow and a ``mixing'' of field operators in the different
intervals, which we describe in the following.

An $n$-interval is the union $\wh E=\bigcup_k \wh I_k$ of $n$ open
intervals with disjoint closure in $S^1$. It is called ``symmetric''
if $\wh I_k$ are the $n$-th roots of an interval $\wh I\subset S^1$. 
We may assume that the point $z=-1$ is not in the closure of $\wh E$, 
i.e., the Cayley preimage $E\subset \RR$ is bounded. Otherwise, we may
first apply a M\"obius transformation. Because M\"obius transformations
intertwine the modular automorphisms, the modular flow would be a
M\"obius conjugate of the following. 

Let the intervals be given by $\wh I_k=(u_k,v_k)\subset S^1$, and define
the function $X$ by     
\be \label{zeta}
X(z):= \prod_k\frac{x-a_k}{x-b_k} \equiv
\prod_k\frac{1+v_k}{1+u_k}\cdot\prod_k\frac{z-u_k}{z-v_k} .\ee
Here, $x,a_k,b_k\in\RR$ are the Cayley preimages of $z,u_k,v_k\in
S^1$.

This function maps each of the intervals monotonously onto $\RR_+$, so
that every $X\in\RR_+$ has exactly $n$ preimages $z_k(X)$, one in each
interval. Then the geometric flow is  
\beno
\delta_t(z_k(X))=z_k(e^{-2\pi t}X).
\ee
The modular automorphism group acts on the Fermi field by
\be \label{modaut}
\sqrt{z'_k(X)}\cdot \sigma_t\big(\wh\psi(z_k)\big) = \sum_j O_{kj}(t,X) \;\sqrt{z'_j(e^{-2\pi t}X)}\cdot\wh\psi(\delta_t(z_j))
\ee
where $z_k'\equiv dz_k(X)/dX$. 
The orthogonal mixing matrix $O(t,X)\in 
SO(n)$ is a cocyle $O(t+s,X)=O(t,X)O(s,e^{-2\pi t}X)$, 
solving the differential 
equation\footnote{This equation is incorrectly displayed in \cite{LMR}
  as $\partial_t O(t)=K(t) O(t)$. The error is due to a change
  of notation between \cite{CH} and \cite{LMR}. Namely, \cite{CH} write
  $\sigma_t(\psi(e^{+2\pi t}X)) = O_{\rm CH}(t,X)\psi(X)$, 
  so that by comparison with \eref{modaut}, $O(t,X)=O_{\rm CH}
  (t,e^{-2\pi t}X)=O_{\rm CH}(-t,X)\inv$. The confusion arises because
  both articles suppress the $X$-dependence of their mixing
  matrices. By \cite{CH}, the differential equation 
  $\partial_t O_{\rm CH}(t,X) =K(e^{+2\pi t}X)O_{\rm CH}(t,X)$ holds. 
  This implies the correct equation \eref{orthdiff}.}
\be \label{orthdiff}
\partial_t O(t,X)=O(t,X)K(e^{-2\pi t}X)
\ee
where
\beno
K(X)_{kj}= 2\pi\frac{\sqrt{z'_k(X)z'_j(X)}}{z_k(X)-z_j(X)}\qquad 
(k\neq j),\,\qquad
K(x)_{kk}=0.
\ee
The solution is a coboundary
\beno
O(t,X)=O(X)^T\cdot O(e^{-2\pi t}X)
\ee
where $O(X)$ is the anti-path-ordered exponential
\be\label{O}
O(X)=\overline P\exp\Big(-\frac1{2\pi}\int_{X_0}^{X} K(X')dX'\Big).
\ee
It follows that the modular mixing is ``diagonalized'' by the
position-dependent orthogonal matrix $O(X)$: Let 
\be \label{diag}
\chi_k(X) := \sum_j O_{kj}(X)\;\sqrt{z'_j(X)}\cdot\wh\psi(z_j(X)). 
\ee
Then \eref{modaut} becomes
\be \label{diagmod}
\sigma_t\big(\chi_k(X)\big) = e^{-\pi t}\chi_k(e^{-2\pi t}X).
\ee
That is, by diagonalizing the modular mixing and reparametrizing
$z_k=z_k(X)$, one recovers the modular automorphisms of $n$
independent Fermi fields in $\RR_+$.

In order to make the connection with Prop.\ 2, let us specialize to
symmetric $n$-intervals. In this case, the related points in $\wh I_k$
are given by $z_k=\omega^kz$, where $\omega=e^{\frac{2\pi i}n}$. It
follows that $-\frac1{2\pi}K(X)dX = \frac Kz dz$, where
$z=e^{i\varphi}$ and $K$ is the constant anti-symmetric matrix with
nondiagonal entries 
\beno
K_{kj}=-\frac {\omega^{\frac{k+j}2}}{\omega^k-\omega^j},
\ee
hence 
\beno
O(X) = \exp \Big(K\int_1^z \frac{dw}w\Big) = z^{K} .
\ee
\medskip

{\bf Lemma:} {\sl The matrix $K$ has integer-spaced spectrum
  $\frac{1-n}2,\dots,\frac{n-1}2$. It is diagonalized by the unitary matrix 
$\frac1{\sqrt n}\,B$, $B_{kj}=\omega^{(\frac12-k)j}$, i.e., $BK=MB$ where $M$ is the
diagonal matrix with entries $m_{kk}=\frac{n+1}2-k$ ($k=1,\dots,n$).
}

\medskip

{\em Proof:} By direct computation (using $\omega^n=1$)
\bea\notag
\sum_{j=1,\dots,n;\, j\neq l} \!\! B_{kj}K_{jl} &=& -\sum_{j=l+1}^{n+l-1} 
\frac {\omega^{(\frac12-k)j}\omega^{\frac{j+l}2}}{\omega^j-\omega^l}
=- \omega^{(\frac12-k)l} \sum_{j=1}^{n-1} 
\frac {\omega^{(\frac12-k)j}
  \omega^{\frac{j+2l}2}}{(\omega^j-1)\omega^l} \\ \notag &\equiv&
- B_{kl}\sum_{j=1}^{n-1} 
\frac {\omega^{(1-k)j}}{(\omega^j-1)} = - B_{kl} \cdot
\frac12 \sum_{j=1}^{n-1} \Big(\frac {\omega^{(1-k)j}}{(\omega^j-1)}  +
\frac {\omega^{(1-k)(n-j)}}{(\omega^{n-j}-1)}\Big) \\ \notag &\equiv&
- B_{kl} \cdot
\frac12 \sum_{j=1}^{n-1} \frac {\omega^{(1-k)j}-\omega^{kj}}{(\omega^j-1)} 
= B_{kl} \cdot
\frac12 \sum_{j=1}^{n-1} \sum_{\nu=1-k}^{k-1}\omega^{j\nu} \\ \notag &=&
B_{kl} \cdot
\frac12 \sum_{\nu=1-k}^{k-1} \big(n\delta_{\nu,0}-1\big) = B_{kl}\cdot
\frac 12(n-2k+1) = m_{kk}\cdot B_{kl}.
\eea
In the first line, we have used the invariance under $j\to j+n$ and
shifted the summation index by $l$. In the second line, we have
symmetrized the sum w.r.t.\ $j\leftrightarrow n-j$. In the third line,
we have cancelled the denominator, and in the last line, we have used 
$\sum_{j=0}^n\omega^{j\nu}=n\delta_{\nu,0}$. \hfill $\square$

\medskip

{\bf Proposition 3:} {\sl The isomorphism given in Prop.\ 2 intertwines the
modular group of $n$ free Fermi fields in an interval $\wh I\subset
S^1$ with the modular group of a single free Fermi field in the
symmetric $n$-interval $\wh E=\sqrt[n]{\wh I}$. }

\medskip

{\em Proof:} By virtue of the lemma, \eref{diag} can be written as 

\be\label{beta-chi}
\sum_j B_{kj} \,\chi_j(X) = \frac{z^{1-k}}n\sum_j\omega^{(1-k)j}
\wh\psi(\omega^jz).
\ee
The right-hand side is precisely $\beta\big(\wh{\phi^{(k)}}(z^n)\big)$
according to Prop.\ 2. We claim that $X(z^n)$ is the composition of the Cayley
transform $z^n\mapsto\frac{z^n-1}{i(z^n+1)}$ that maps $\wh I$ onto
$I$, with a M\"obius transformation that maps $I$ onto $\RR_+$. Both 
M\"obius transformations and the isomorphism $\beta$ preserve the vacuum
state and hence intertwine the respective modular
automorphisms. Knowing the modular automorphisms of
$\wh{\phi^{(k)}}(z^n)$ for $z^n\in \wh I$, it follows that the modular
automorphisms of $\chi_j(X)$ are given by the dilations $X\to e^{-2\pi t}X$ 
on $\RR_+$. This explains the modular automorphisms \eref{diagmod} and
hence also \eref{modaut}. It only remains to verify the claim that the
function $X$ given by \eref{zeta} is indeed of the said form. But, in
the symmetric case one may use identities like
$\prod_k(z-w_k)=z^n-w^n$ to find 
\be \notag
X=-\frac{(-1)^n-v^n}{(-1)^n-u^n}\cdot\frac{z^n-u^n}{z^n-v^n}
\ee
which is indeed a M\"obius transform of $\frac{z^n-1}{i(z^n+1)}$. \hfill $\square$

\medskip

It should be noted that by what has been said, \eref{diag} 
diagonalizes the modular flow only inside the 2-interval 
($z_k\in\wh I_k$, $z^n\in\wh I$), whereas the isomorphism $\beta$
manifestly extends to the entire circle. This reflects the fact that
the graded commutant (denoted by $\cdot^c$) of the Fermi field algebra
in an $n$-interval equals the Fermi field algebra in the complement of
the $n$-interval: 
\beno 
\pi_0\big(\CAR(\wh E)\big)^c = \pi_0\big(\CAR(S^1\setminus \wh E)\big)'',\quad \pi_0\big(\CAR^n(\wh I)\big)^c =
\pi_0\big(\CAR^n(S^1\setminus \wh I)\big)''.
\ee
By Modular Theory, the modular flow of the graded commutant is
given by the inverse of the modular flow of the original algebra. 
Thus, the same orthogonal matrix $O(X)$ also diagonalizes the modular
flow on the complements.  

Finally, our result answers a question raised by M. Bischoff and
Y. Tanimoto: If $\wh E$ is a symmetric $n$-interval, and $\wh F\subset
\wh E$ is a symmetric subinterval with common upper limits of the
intervals, let $M:=\pi_0\big(\CAR(\wh E)\big)''$ and
$N:=\pi_0\big(\CAR(\wh F)\big)''$. Then, the data
$(N_0,N_1,N_2,\Omega)$ with $N_0=M^c,N_1=N,N_2=N^c\cap M$ form a 
``$+$half-sided modular factorization'' in the sense of 
\cite[Thm.\ 1.2]{GLW} (actually, a graded generalization thereof: one
has $N_i\subset N_{i+1\mod 3}^c$, and the modular automorphisms
$\sigma_{-t}$ of the larger algebra map the smaller algebra into
itself for $t\geq 0$). By 
the main theorem of \cite{GLW}, they therefore define a (graded local)
conformal quantum field theory such that its local algebras are given
by $A(\RR_+)=M$, $A(\RR_++1)=N$, and the M\"obius transformations are
generated by the modular groups of the three algebras.   

\medskip
{\bf Corollary 5:} {\sl The conformal QFT obtained from the graded $+$-hsm
factorization $(N,M^c,N^c\cap M,\Omega)$ is isomorphic to the free
Fermi theory $\CAR^n(S^1)$ of $n$ free Fermions.  }

\medskip

Namely, the isomorphism is just given by $\beta\inv$, which maps $M$ onto
$\pi_0\big(\CAR^n(\wh I)\big)''$ and $N$ onto $\pi_0\big(\CAR^n(\wh J)\big)''$ 
(where $\wh E=\sqrt[n]{\wh I}$, $\wh F=\sqrt[n]{\wh J}$), combined
with a M\"obius transformation which maps $I$ onto $\RR_+$ and $J$ onto
$\RR_++1$. Both isomorphisms intertwine the respective vacuum modular groups.

\section{The Ramond case}
\setcounter{equation}{0}

The real free Fermi field possesses another faithful representation
of positive energy: the Ramond representation induced (by the GNS
construction) from the quasifree ground state with 2-point function 
\beno
\omega_R\big(\psi(x)\psi(y)\big)=\frac{1+xy}{\sqrt{1+x^2}\sqrt{1+y^2}}\cdot
\frac{-i}{x-y-i\eps}.
\ee
In the compact picture on $S^1\setminus\{-1\}$, this is 
\beno
\omega_R\big(\wh\psi(z)\wh\psi(w)\big)=\frac{z+w}{2\sqrt{zw}}\cdot\frac1{z-w}.
\ee
Obviously, in the Ramond representation, $\pi_R\big(\wh\psi(z)\big)$
has a cut at $z=-1$ and extends anti-periodically to $S^1$. It is
therefore convenient to introduce the Ramond field
\beno
\psi_R(z):= \sqrt{z}\cdot\pi_R\big(\wh\psi(z)\big)
\ee 
which extends periodically to $S^1$ with Fourier representation 
\beno
\psi_R(z)=\sum_{n\in\ZZ}\psi_{R,n} z^{-n},
\ee
where $\psi_{R,n}^*=\psi_{R,-n}$ satisfy the CAR
$\{\psi_{R,n},\psi_{R,m}\}=\delta_{n+m,0}$. In particular, this field has a zero
mode with $2\psi_{R,0}{}^2=1$. 

Because we are going to consider the field both in the vacuum and in
the Ramond representation, we shall below write the field in the
vacuum representation as 
$\psi_0(z):=\pi_0\big(\wh\psi(z)\big)$, and write no subscript when
the field is understood algebraically. 

The modular theory of the Ramond field is not known. We therefore lack
a rationale to expect a similar isomorphism as in Prop.\ 1, which
would explain a modular mixing as in Prop.\ 3. 

Nevertheless, one can prove along the same lines as in Prop.\ 1 the
following 

\medskip

{\bf Proposition 4:} {\sl The linear map 
\bea\notag
\beta_R:\psi_R^{(1)}(z^2)\equiv\psi_R(z^2)\otimes^t \eins\mapsto
\frac1{2}\big(\psi_R(z)+\psi_R(-z)\big),
\\ \label{betaR}
\psi_0^{(2)}(z^2)\equiv\eins\otimes^t\psi_0(z^2)\mapsto \frac1{2z}\big(\psi_R(z)-\psi_R(-z)\big)
\eea
for $z\in S^1$, induces an isomorphism of CAR algebras
$\beta:\pi_R\big(\CAR(S^1)\big)\otimes^t \pi_0\big(\CAR(S^1)\big)\to \pi_R\big(\CAR(S^1)\big)$, such that 
$\omega_R\circ\beta_R = \omega_R\otimes\omega_0$.} 

\medskip 

Namely, in terms of the Fourier modes, the right-hand-sides equal,
respectively, $\sum_{n\in\ZZ}\psi_{R,2n}z^{-2n}$ and
$\sum_{n\in\ZZ}\psi_{R,2n+1}z^{-2n-2}$, from which the correct commutation
relations and two-point functions follow. 

\medskip

The fact that the current \eref{standard}
$j(x)=i\wick{\psi^{(1)}(x)\psi^{(2)}(x)}$ satisfies the CCR \eref{CCR} 
is purely algebraic, and therefore independent of the
representation. 
Taking $\psi^{(1)}$ in the Ramond representation and $\psi^{(2)}$ in
the vacuum representation, the isomorphism \eref{betaR} embeds the 
resulting current into the Ramond algebra
$\pi_R\big(\CAR(S^1)\big)$. The result is (in the compact picture)
\be\label{currR}
\beta_R\big(\wh j(z^2)\big) = \frac1{2iz^2}\cdot\wick{\psi_R(z)\psi_R(-z)}_{\!R}\ee
where $\wick{\cdot}_{\!R}$ stands for the Wick product defined with the
subtraction of the Ramond expectation value (which is zero in the
case at hand). The right-hand side is the formula given in
\cite{A1}. The new aspect here is that it arises from an underlying
isomorphism of CAR algebras.  

The embedded current \eref{currR} changes sign under $z\mapsto-z$,
hence is anti-periodic in the variable $z^2$. In other words, we find the
current in a representation that extends anti-periodically to the
circle. 
Indeed, this representation of the current is known as the ``twisted'' 
representation \cite{A1,Fr} obtained from the quasifree state with
2-point function given by \eref{twisted}. This can be established by
evaluating the embedded current in the Ramond state. We find by direct
computation 
\beno
\omega_R\big(\beta_R(\wh j(z^2))\big) = 0,\qquad
\omega_R\big(\beta_R(\wh j(z^2))\beta_R(\wh j(w^2))\big)
= \frac{z^2+w^2}{2zw(z^2-w^2)^2}.
\ee
The state $\omega_R\circ\beta_R$ restricted to the current
is again quasifree. Thus, the embedding of the
current via the isomorphism \eref{betaR} produces the twisted
representation of the CCR algebra: 
\beno
\omega_R\circ\beta_R = \omega_t \, .
\ee

Finally, we have computed the stress-energy tensor by using the
bosonic formula $T^{\rm curr}(x)=\frac1{4\pi}\wick{j(x)^2}$. 
We find the same formula as \eref{betaT}:
\beno
\beta_R\big(T^{\rm curr}(z^2)\big) = \frac1{4z^2}\Big(\pi_R\big(T^{c=\frac12}(z)\big)+\pi_R\big(T^{c=\frac12}(-z)\big)\Big) + \frac{1}{64\pi z^4}\,.
\ee
(In deriving this equation, care must be taken of the fact that due to
the different subtractions, one has 
$\wick{\wh\psi(z)\partial_z\wh\psi(z)}_{\!R} =
\wick{\wh\psi(z)\partial_z\wh\psi(z)}_{\!0} + \frac 1{8z^2}$.) 
Since $L^{c=\frac12}_0$ has eigenvalue $h_R=\frac1{16}$ in the Ramond state, we
conclude that $\omega_R(\beta_R(L^{\rm curr}_0))=\frac 1{16}$, in agreement
with the ground state energy in the twisted sector of the current.  

\section{Conclusion}
We have described a multilocal isomorphism of the CAR algebra of
a real chiral Fermi field with the CAR algebra of any number $n$ of
such fields. The isomorphism preserves the vacuum state. This has two
consequences: 

(1) The isomorphism extends to Wick products, and
therefore allows to ``embed'' the generators of local symmetries
(gauge transformations or diffeomorphisms) of one theory into the
other. We have explicitly displayed the ``embedded'' symmetries and their
multilocal features in the case $n=2$. A characteristic feature is
that the gauge transformations of the complex Fermi field, embedded
into the algebra of the real Fermi field, generate a mixing of Fermi
fields at different points; and that the embedded diffeomorphisms
consist of a mixing on top of a geometric transformation. 

(2) A similar mixing was found earlier to occur in the modular
automorphisms of the chiral Fermi algebra in multiple intervals. We
have demonstrated that this modular mixing is a special case of the
mixing going with the embedded diffeomorphisms, when the latter arise as
modular automorphisms for a single interval, i.e., certain one-parameter
groups of M\"obius transformations. However, as stated in (1), the
embedded diffeomorphisms are not restricted to M\"obius transformations,
and, in view of the fact that the vacuum state is not diffeomorphism
invariant, are not necessarily related to modular theory.

Finally, we observed that a similar isomorphism exists in the Ramond
representation, and we have computed its restriction to the subalgebra
of the current and the stress-energy tensor.

\bigskip

{\bf Acknowledgements:}
This work was supported in part by the German Research Foundation
(Deutsche Forschungsgemeinschaft (DFG)) through the Institutional
Strategy of the University of  G\"ottingen, and by the DFG Research
Training School 1493 ``Mathematical Structures in Modern Quantum
Physics''. KHR is grateful to I.~Anguelova for interesting
discussions, which triggered this work, and to Y.~Tanimoto and
M.~Bischoff, who raised interesting questions.

\end{document}